\documentclass[
 amsmath,amssymb,
 aps,onecolumn]{revtex4-2}
\usepackage{graphicx} 
\usepackage{dcolumn}
\usepackage{bm}
\usepackage{amsmath}
\usepackage{epsfig}
\usepackage{color}

\makeatletter
\def\maketitle{
\@author@finish
\title@column\titleblock@produce
\suppressfloats[t]}
\makeatother

\begin{document}

\title{Dynamic redundancy as a mechanism to optimize collective random searches}

\author{Daniel Campos and Vicen\c{c} M\' endez}

\address{Grup de F\' isica Estad\' istica, Departament de F\' isica. Facultat de Ci\`encies), Universitat Aut\`onoma de Barcelona, 08193 Bellaterra (Barcelona), Spain}

\begin{abstract}
   We explore the case of a group of random walkers looking for a target randomly located in space, such that the number of walkers is not constant but new ones can join the search, or those that are active can abandon it, with constant rates $r_b$ and $r_d$, respectively. Exact analytical solutions are provided both for the fastest-first-passage time and for the collective search time required to reach the target, in the seminal case of Brownian walkers with $r_d=0$. We prove that even for such a simplified situation there exists an optimal rate $r_b$ at which walkers should join the search to minimize the collective effort required to reach the target. We discuss how these results open a new line to understand the optimal regulation of cooperative random searches, e.g. for the case of biological foraging in social species.
\end{abstract}

\maketitle

\section{Introduction}

In the last years, there has been an increasing interest about the statistical properties of the so-called \textit{redundancy principle} and the role it may play in biochemical and other biological processes \cite{tononi1999,min2011,schulman2015,hebets2016,farnsworth2017,laruson2020,langella2021,goldford2022,hunter2022,ghanbari2023}. This principle is based on the notion that \textit{'many copies of a single object such as molecules, cells, etc... is not a waste, but it has a specific function in living systems'} \cite{schuss2019}. This has been explored in particular as a mechanism to explain how chemical activation rates at the cellular level can be regulated or facilitated by the presence/creation of a very large (redundant) number of activators. When this happens, the statistical properties of the activation process will be governed by the fastest activator to reach the receptor, so providing an attractive link to extreme value statistics \cite{schuss2019,redner2019,rusakov2019,sokolov2019,tamm2019,paquin2022,lawley2023}.

There is actually a relatively rich literature, which extends over the last 40 years, about the problem of the fastest-first-passage time (FFPT) of a group of $N$ random searchers/particles through a target located at a distance $\mathbf{x}_0$ from their initial position (see, e.g., \cite{lindenberg1980,weiss1983,krapivsky1996,yuste1996,yuste2000,yuste2001,rojo2010,mejia2011,meerson2014,meerson2014b,meerson2015,ro2017,grebenkov2017,lawley2020,madrid2020,basnayake2020,toste2022,grebenkov2022,linn2022}). For the case of Brownian particles in 1d infinite domains, for example, it is known that the mean fastest-first-passage time (MFFPT) becomes finite for $N \geq 3$ \cite{rojo2010} (remember that the mean first-passage time for a single-walker, i.e. $N=1$, diverges), and it has a leading term $\langle T \rangle = \mathbf{x}_0^2/ (4D \log{N})$ for $N \rightarrow \infty$ \cite{weiss1983,yuste2001}, with $D$ the diffusion coefficient of the particles. For two and higher dimensions, instead, the MFFPT remains infinite for any finite value of  $N$ \cite{lindenberg1980,rojo2010}. Also, some approximations and series expansions have been proposed for the corresponding FFPT distribution \cite{yuste2001,lawley2020b,lawley2020c}, but exact results are extremely difficult to obtain.

The fact that redundant walkers/trajectories can modify the first-passage statistics of the process, or they can even turn (at least in 1d) the MFFPT finite, is also reminiscent of the case of random trajectories with stochastic resetting and other stochastic processes with spatial memory \cite{evans2011,maso2019,evans2020,gupta2022,nagar2023}. There, the possibility to restart the trajectory anew gives the walker the option to rectify strong departures from the target location. Stochastic resetting has also been studied for the $N$-walkers case, but this has been limited to simplified situations or approximated results (some examples are those in \cite{bhat2016,evans2020} and the references therein).

In this Letter we present a more general version of these problems that (as we shall show) still allows an exact analytical treatment in order to explore how the idea of redundancy applies to cooperative random searches. We introduce the notion of \textit{dynamic redundancy}, which corresponds to the case where the number of searchers $N$ is not fixed but changes with time, so that each searcher can join the foraging process and/or abandon it later. This is inspired in the case of eusocial biological species, like honeybees or ants, where the rate at which new individuals emerge out of the nest for foraging is regulated internally by the colony (according to some communication rules and other mechanisms). Surprisingly, this case of a non-constant number of searchers has received relatively low attention in the literature of random searches; a few exceptions are those works in which walkers are assumed to become inactivated/vanished after some time (see, e.g, \cite{grebenkov2017,grebenkov2022,lawley2023}), or they can aggregate \cite{choi2021}. Instead, the case where new searchers can join the process at $t>0$ apparently has not been considered to date. Therefore, in the following we introduce formally the problem, and then study some seminal cases to illustrate its interest.

 \begin{figure*}[t]
	\centering
		\includegraphics[width=0.6\textwidth]{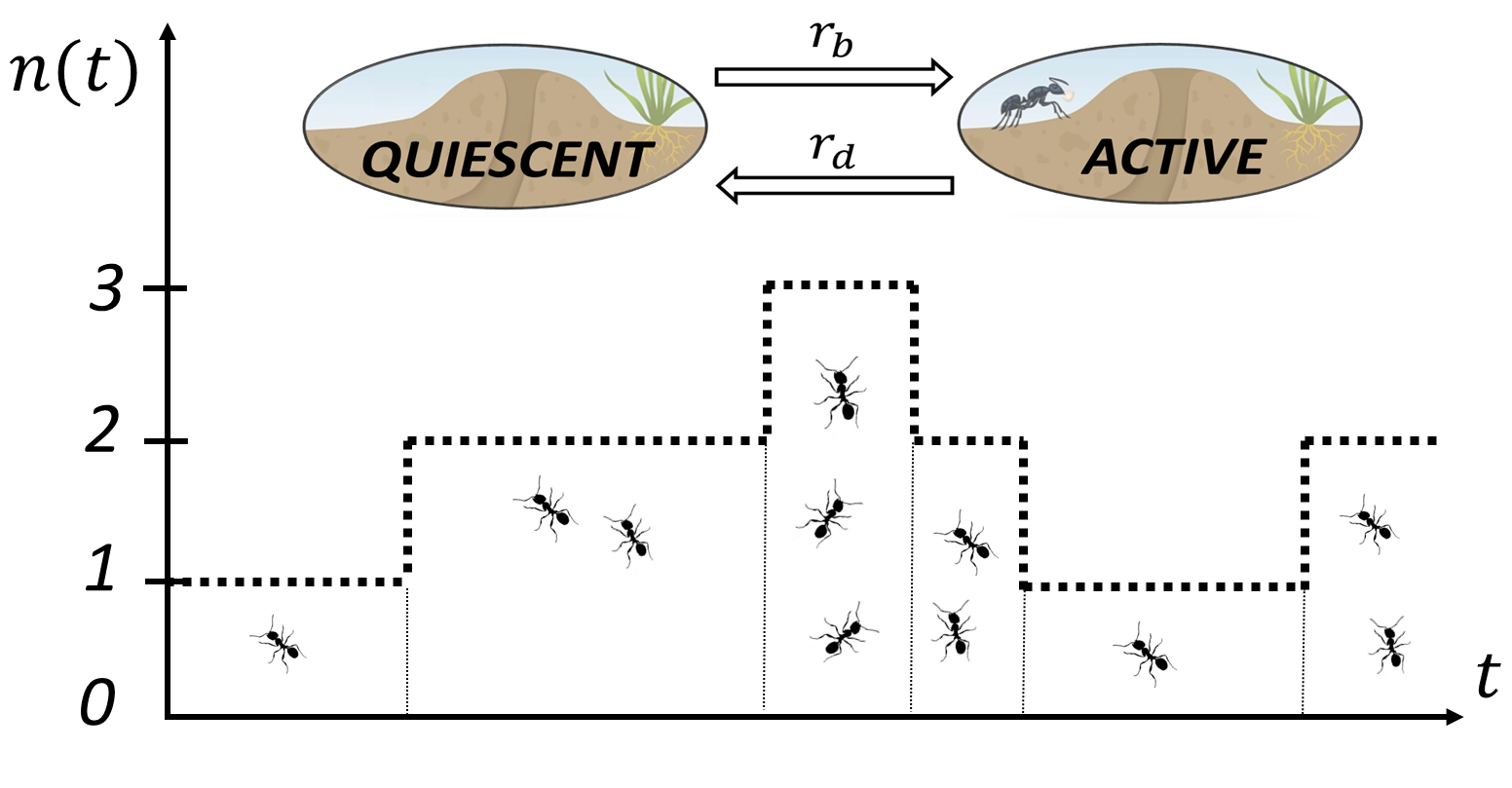}
		\caption{Schematic representation of the dynamic redundancy problem, where the number of walkers searching for a random target, $n(t)$, evolves in time according to a birth-death dynamics between a \textit{quiescent} (where the individual stays quiet in the nest) and an \textit{active} (random-walk) state.}
	\label{fig1}
\end{figure*}

\textbf{Dynamic redundancy problem.} We consider a group of walkers which are assumed to be initially at the same position $\mathbf{x}_0$ (which represents their nest) in a \textit{quiescent} state, so they cannot move from that location. Then, we define $t=0$ as the time at which one walker suddenly switches to the \textit{active} state, which means that it starts a random walk trajectory from $\mathbf{x}_0$. From then on, the number of walkers in the \textit{active} state, $n(t)$, is assumed to follow a birth-death stochastic dynamics such that (i) a new \textit{active} individual can emerge from the nest at some random time $t>0$, and (ii) any of the \textit{active} walkers can, with a certain probability, become \textit{quiescent} again (so its position is instantaneously reset to $\mathbf{x}_0$). For simplicity, we will assume that the transition between the two states is Markovian, so the birth-death process is governed by constant rates $r_b$ and $r_d$ (see Fig. \ref{fig1}).

Our specific aim in the present work is twofold. First, we want to understand the properties of the FFPT (denoted by $T$) that it takes for any walker of the group to reach a target located at $\mathbf{x}=0$, as a function of the birth-death dynamics described above and the distance to the nest $\vert \mathbf{x}_0 \vert$. Second, we want to check whether there is an optimal shape of $n(t)$ (that is, optimal rates $r_b$ and $r_d$) that can be used to reach the target with a minimum effort. For the latter, we introduce the concept of collective search time (CST) as

\begin{equation}
    T_c(t)=\int_{0}^{t} n(t') dt'.
\label{intensity}
\end{equation}

This can be interpreted as the sum of the individual search times spent by every \textit{active} walker up to time $t$. Then, $T_c(t=T)$ represents the CST up to the FFPT. Formally, $T_c(T)$ represents a first-passage functional subordinated to the birth-death process $n(t)$ \cite{majumdar2020,singh2021}. Note that $T$ itself depends explicitly on $n(t)$, so this makes the computation of $T_c(T)$ nontrivial. Minimizing the CST requires reducing $T$ but without involving too many walkers in the process, so a proper balance between keeping both $T$ and $n(t)$ small is required; here we will focus on understanding the specific statistical properties of that balance.

\textbf{Connection to previous search problems}. The dynamic redundancy problem here presented can be reduced to some previously studied cases of interest. Obviously, the case $r_b=r_m=0$ corresponds to the classical first-passage problem for a single walker, while the case $r_b=0$, $r_m \neq 0$ would correspond to the case of mortal random walkers \cite{abad2012,yuste2013,campos2015,meerson2015,lawley2023,radice2023}. Also, the situation where $r_b \ll r_d$ can be mapped into the case of stochastic resetting, with $r_d$ representing then the reset rate. In this case, the birth-death process will take most of the time the value $n(t)=0$, and from time to time a single \textit{active} walker will emerge from the nest ($n(t)>1$ being extremely unlikely). Since the \textit{quiescent} periods where $n(t)=0$ do not contribute to the integral in (\ref{intensity}), the statistics of the CST $T_c(T)$ will then correspond to that of the first-passage time for the stochastic resetting problem. This means, for instance, that its average value will tend to $ \langle T_c \rangle  = \left(  e^{x_0\sqrt{r_d/D} } -1 \right) /r_d$ for the case of 1d Brownian walkers with a diffusion coefficient $D$ \cite{evans2011}.

\textbf{Exact solution for $r_d =0$.} In the following we will focus in the case where individual search trajectories are never terminated until the target is found by the group, so $r_d=0$. Then, the stochastic process $n(t)$ becomes strictly increasing. As we shall show, this particular case suffices to investigate the aforementioned balance necessary for minimizing the CST on average.

We will define $f(t)$ as the probability distribution for the FFPT, and the corresponding survival probability of the target is denoted as $S(t)=\int_t^{\infty} f(t')dt'$. For convenience, we will separate all the realizations of the stochastic process that lead to the target detection at time $t$ into those which correspond to $n(T)=1$, $n(T)=2$, and so on. So that, we define the survival probability for the case $n(T)=N$ as $S_N(t)$, so this corresponds to the probability that the target has not been reached yet at time $t$, provided that there are $N$ \textit{active} walkers by that time. Note that the expression for $N=1$ will satisfy $S_1(t)=S_{sw}(t)e^{-r_b t}$, where $S_{sw}(t)$ represents the survival probability for the classical case of a single walker (it is, for $r_b=r_d=0$), and the exponential factor $e^{-r_b t}$ is the probability that no additional \textit{active} walkers have emerged in the interval $(0,t)$.

Through these definitions, we can then write the recurrence relation (see Supplementary Information for details)

\begin{equation}
    S_N (t)= r_b e^{-r_b t} \int_0^t S_{sw}(t) S_{N-1} (t-t_1) dt_1,
    \label{survivalrec}
\end{equation}
where the integration variable $t_1$ represents the random time at which the second walker becomes \textit{active}, it is, the time at which $n(t)=1$ switches to $n(t)=2$. So that, in (\ref{survivalrec}) the term $S_1(t)=S_{sw}(t) e^{-r_b t}$ is being multiplied by $r_b \int_0^t S_{N-1}(t-t_1) dt_1$, where the latter provides the survival probability for the remaining $N-1$ \textit{active} walkers over all possible values of $t_1$. The previous relation allows us to find recurrently the general expression for $S_N(t)$. Using this procedure a simplified expression is obtained in the form

\begin{equation}
    S_N(t)= e^{-r_b t} S_{sw}(t) \frac{\left[ r_b g(t) \right] ^{N-1}}{(N-1)!}
    \label{survivalN}
\end{equation}
where we have introduced $g(t) \equiv \int_0^t S_{sw}(t') dt'$. Now, the general expression for the survival probability in our dynamic redundancy problem is easily derived as

\begin{equation}
    S(t) = \sum_{N=1}^{\infty} S_N(t) = S_{sw}(t) e^{-r_b (t-g(t))},
    \label{survivaldef}
\end{equation}
and the corresponding first-passage distribution is simply the time derivative (with a minus sign) of $S(t)$; in the Supplementary Information file we provide the whole expressions for this as well as additional discussions.

For computing $T_c$ one can follow a similar procedure to that above. So, we will define $\overline{T_c^{(N)}}(t)$ as the contribution to the mean collective search time (MCST) that comes from all the realizations of the search process for which $n(T)=N$. Hence, the overall MCST (computed over all possible values of $N$ and all possible values of the FFPT) is simply

\begin{equation}
    \langle T_c \rangle = \int_0^{\infty} dT \overline{ T_c } (T) = \int_0^{\infty} dT \sum_{N=1}^{\infty} \overline{T_c^{(N)}}(T). 
    \label{meanI}
\end{equation}

It is possible to check that the expressions for $\overline{T_c^{(N)}}(T)$ also satisfy a recurrent relation which allows us to write them in terms of $S^{(N-1)}(t)$ and $\overline{T_c^{(N-1)}}(t)$, though in this case the interpretation of this recurrence is much less straightforward . Using that recurrent relation, one can derive again the expression for the integrand in the MCST (\ref{meanI}) (all technical details and discussions are provided in the Supplementary Information file):
\begin{eqnarray}
   \overline{ T_c } (T) &=&e^{-r_{b}\left[T-g(T)\right]}\left\{ r_{b}S_{sw}(T)\left(T+g(T)-2TS_{sw}(T)\right)\right.\nonumber\\
    &+&\left. r_{b}^{2}h(T)S_{sw}(T)\left(1-S_{sw}(T)\right)\right.\nonumber\\
    &+&\left. f_{sw}(T)\left(T+r_{b}h(T)\right)\right\} 
    \label{intensitydef}
\end{eqnarray}
with $h(t) \equiv \int_0^t t'S_{sw}(t') dt'$, and where we define $f_{sw}(t)=-dS_{sw}/dt$ as the first-passage time distribution for the classical (single-walker) case.

\textbf{Brownian walkers in 1d.} While we have now some exact expressions for the survival probability and for the collective search times, it is not trivial in general to compute their average properties. Let us consider the seminal case in which all \textit{active} walkers behave as Brownian particles moving in a 1d domain with diffusion coefficient $D$. For this, it is well-known that $S_{sw}(t) = \mathrm{erf} \left( \frac{x_0}{2 \sqrt{D t}} \right)$, with $\mathrm{erf}(\cdot)$ representing the error function \cite{feller1968}. However, introducing this expression into (\ref{survivaldef}) or (\ref{intensitydef}) does not allow to reach an exact expression neither for the mean-first passage time $\langle T \rangle = \int_0^{\infty} S(t) dt$ nor for the MCST (\ref{meanI}). So, we will take the two timescales driving the search process, i.e. the diffusive scale $x_0^2/4D$ and the birth scale $r_b^{-1}$, and will focus on the behavior for the limit cases where the dimensionless parameter $\chi\equiv x_{0}^{2}r_{b}/4D$ becomes either small or large.

Integrating by parts and using (\ref{survivaldef}), the MFFPT $\langle T \rangle$ can be written in the scaling form
\begin{eqnarray}
      \left\langle T\right\rangle =\frac{x_{0}^{2}}{2D}\phi(\chi)
      \label{mfpt2}
\end{eqnarray}
where $\phi (\chi)$ is defined in the Supplementary Information (SI) file. Using this scaling, the leading term of $\langle T \rangle$ can be properly computed in the limits $\chi \ll 1$ and $\chi \gg 1$ (the corresponding derivation is also provided in the SI file). This leads to

\begin{equation}
\left\langle T\right\rangle \simeq\left\{ \begin{array}{cc}
\frac{x_{0}}{\sqrt{Dr_{b}}} ,& r_{b}\ll4D/x_{0}^{2}\\
\frac{x_{0}^{2}}{4D \ln\left( \sqrt{\frac{2}{243 \pi}} \frac{x_{0}^{2}r_{b}}{D}\right)} ,& r_{b}\gg4D/x_{0}^{2}.
\end{array}\right.
\label{mfptlimits}
\end{equation}

Similarly, the CST also admits a scaling form
\begin{eqnarray}
    \langle T_{c}\rangle =\frac{x_{0}^{2}}{2D}\phi_c(\chi),
    \label{mtc}
\end{eqnarray}
where $\phi_c(\chi)$ is also defined in the SI file. Its leading order behavior in the two limits $\chi \ll 1$ and $\chi \gg 1$ above reads

\begin{equation}
\left\langle T_c \right\rangle \sim \left\{ \begin{array}{cc}
\frac{x_{0}}{\sqrt{Dr_{b}}} ,& r_{b}\ll4D/x_{0}^{2}\\
\frac{\chi}{\left( \ln{\chi} \right)^2} ,& r_{b}\gg4D/x_{0}^{2}.
\end{array} \right.
\label{ctlimits}
\end{equation}

\begin{figure*}[t]
	\centering
		\includegraphics[width=0.6\textwidth]{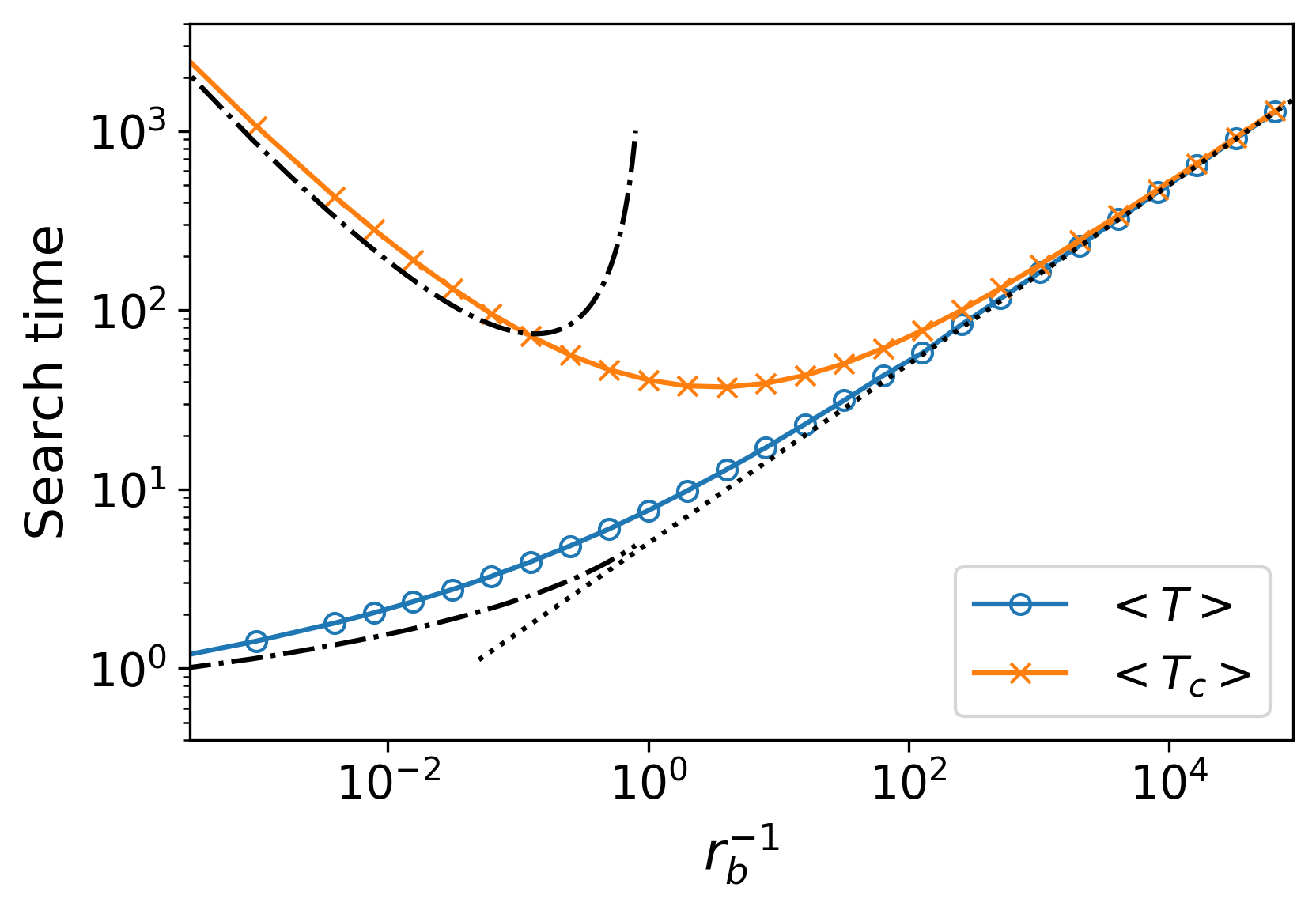}
		\caption{MFFPT (in blue) and the MCST (in orange) as a function of the birth timescale $r_b^{-1}$. Solid lines correspond to the exact values computed from numerical integration, and symbols correspond to random-walk simulations. Dotted and dotted-dashed lines represent the analytical asymptotic behavior predicted in (\ref{mfptlimits}) and (\ref{ctlimits}). All the results shown correspond to the case $x_0=5$, $D=1$.}
	\label{fig2}
\end{figure*}

Eqs. (\ref{mfptlimits}) and (\ref{ctlimits}), together with the general expressions (\ref{survivaldef}) and (\ref{intensitydef}) from which they are derived, represent the main results of our work. To visualize these results, in Figure \ref{fig2} we plot both the MFFPT and the MCST as a function of the birth scale. There, the solid lines correspond to the exact values of these two quantities obtained from numerical integration of our exact expressions above, and the symbols correspond to random-walk simulations that we have carried out (averaging over $10^4$-$10^5$ realizations of the collective search process) to confirm the validity of our derivations. From the plots obtained, we see how the asymptotic results indicated above for the limits $\chi \gg 1$ and $\chi \ll 1$ (represented through dotted and dashed-dotted lines, respectively) are recovered in the appropriate regimes.

It is interesting to note that in the regime $\chi \ll 1$ one finds $\langle T \rangle \approx  \langle T_c \rangle$. This is because for this regime, in most of the realizations the target will be reached by the first \textit{active} walker before the second one becomes \textit{active}. There is only a relatively small probability that the first \textit{active} walker departs from the target as time goes by and misses the target. For  $r_b=0$ this would lead to the divergence of both $\langle T \rangle$ and $\langle T_c \rangle$, while for $r_b$ small  the target is found in a finite time thanks to the new \textit{active} walkers eventually appearing. From the properties of a diffusive flux to an absorbing boundary in a 1d semiinfinite domain (see, e.g, \cite{redner2001}), we know that the fraction of the trajectories that lead the first searcher far away from the target decreases as $\sim t^{-1/2}$, so this explains the scaling $\langle T \rangle \sim 1/\sqrt{r_b}$ obtained in (\ref{mfptlimits}).

The opposite situation, $\chi \gg 1$, would correspond to the case where a very large number of \textit{active} walkers emerge in a time much smaller than the diffusive scale $x_0^2/D$. Such situation is then reminiscent of the case of a set of $N$ walkers searching from the target starting from the same initial position. As mentioned above, it is known for this case that in the limit $N \rightarrow \infty$ the leading order term of the MFFPT satisfies $\langle T \rangle \sim (\log{N})^{-1}$. Since on average the number of \textit{active walkers} appearing will be proportional to $r_b$, this explains the scaling $\langle T \rangle \sim (\log{r_b})^{-1}$ we observe. Also, for this regime where the process is dominated by large values of $N$ we can approximate $\langle T_c \rangle = \langle \frac{1}{2}  N T \rangle \sim \langle N \rangle \langle T \rangle$, where $\langle N \rangle$ is the mean number of \textit{active} walkers we have at $t=T$. Note that the factor $\frac{1}{2}$ appears from the fact that any of these walkers will be \textit{active} on average during a time $T/2$. Then, we note that on average $\langle N \rangle  = r_b \langle T \rangle$ is satisfied. So that, we realize that $\langle T_c \rangle \approx \frac{1}{2} r_b \langle T \rangle^2$, and combining this with the scaling above for $\langle T \rangle$ (see Eq. \ref{mfptlimits}) we finally obtain $\langle T_c \rangle \sim \chi /\left( \ln{\chi} \right) ^2$, which confirms intuitively the formal result in Eq. (\ref{ctlimits}).

Anyway, the most relevant feature in Figure 2 is the fact that while $\langle T \rangle$ decays monotonically with $r_b$, there is instead an optimum rate which minimizes $\langle T_c \rangle$, so providing a criteria to optimize these collective searches. To understand better this optimization mechanism, in Figure 3 we show again the behavior of $\langle T_c \rangle$ as a function of $r_b^{-1}$, but now for different values of $x_0^2/D$. This allows us to see how the dynamic redundancy problem can be optimized as a function of the distance from the nest to the target. If we denote by $(r_b^{*},\langle T_c \rangle ^{*})$ the coordinates of the point at which the MCST reaches its minimum, we observe that (i) $r_b^*$ increases linearly with the diffusive timescale, such that $r_b^{*} \approx 7.536 D/x_0^2$, and (ii) the minimum value of the MCST scales as $\langle T_c \rangle^{*} \approx 1.489 x_0^2/D$, so this means that $\langle T_c \rangle^{*} \sim (r_b^{*})^{-1}$ (additional justification for these relations is provided in the Supplementary Information file).

 \begin{figure*}[t]
	\centering
		\includegraphics[width=0.6\textwidth]{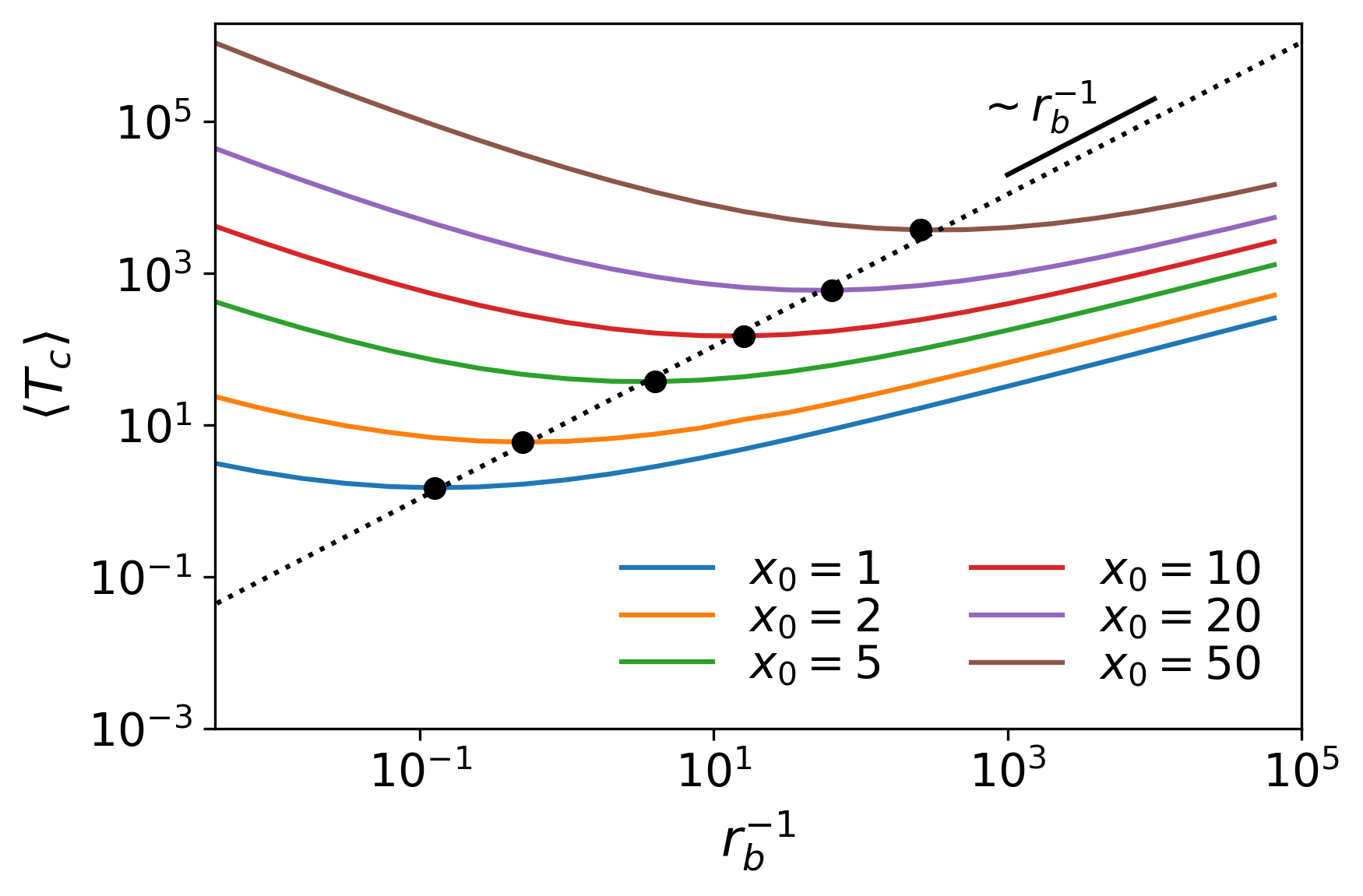}
		\caption{MCST as a function of $r_b^{-1}$ for different values of the distance between nest and target $x_0$. Full symbols denote the minimum search intensity for each case and the dotted line provides a linear fit between them to confirm the scaling $\langle T_c \rangle ^{*} \sim (r_b^{*})^{-1}$. In all cases $D=1$ is used.}
	\label{fig3}
\end{figure*}

\begin{figure*}[t]
	\centering
		\includegraphics[width=0.6\textwidth]{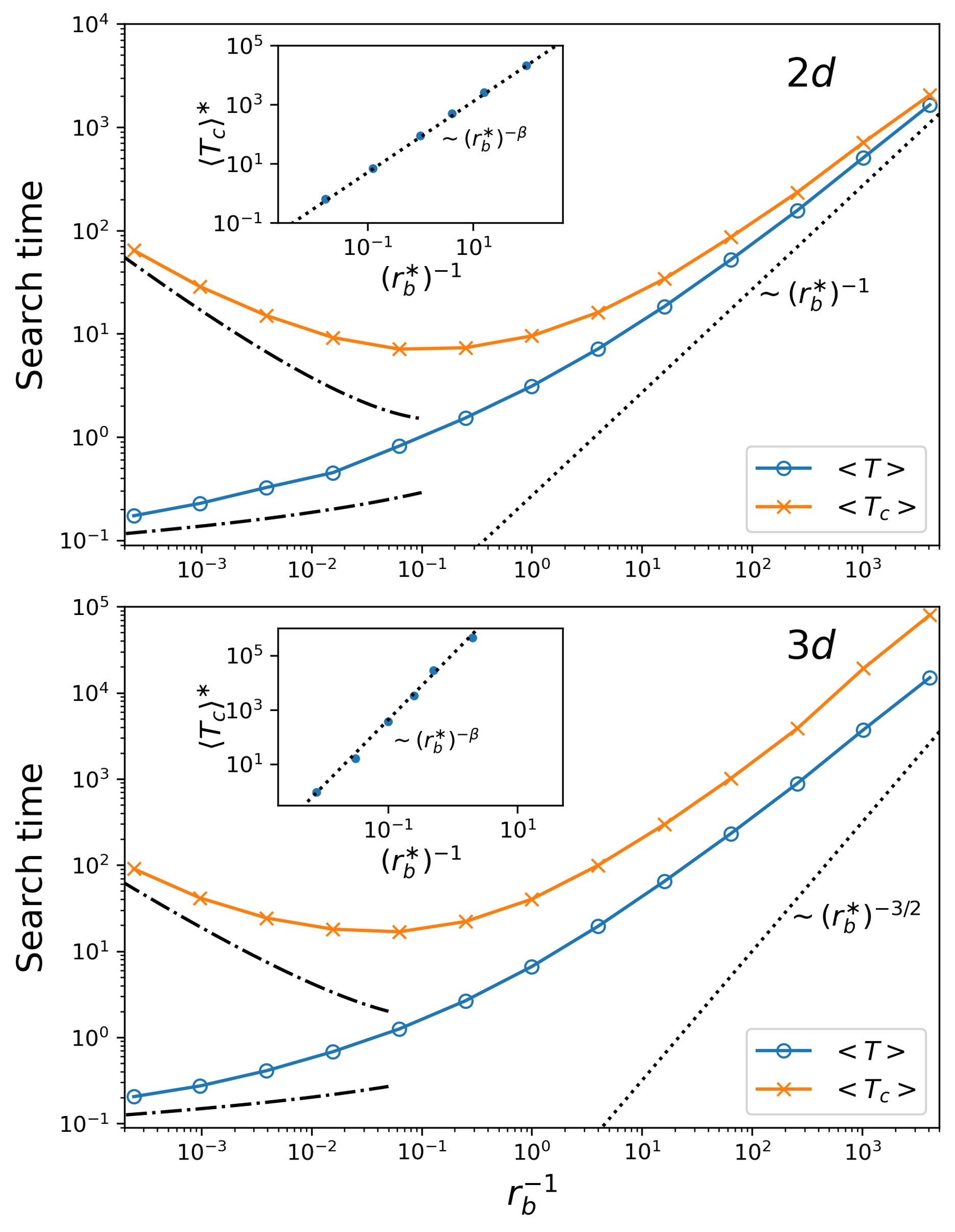}
		\caption{MFFPT (in blue) and the MCST (in orange) for two and three dimensions (upper and lower panels, respectively) as a function of the birth timescale $r_b^{-1}$. Symbols correspond to random-walk simulations, and the solid lines in this case are only provided as a visual guide. Dotted and dashed-dotted lines are the limit behavior expected for $\chi \ll 1$ and $\chi \gg 1$, respectively. All results shown correspond to the case $x_0=2$, $D=1$. (Insets): minimum of the MCSTe when computed for different values of $x_0$, as in Fig. \ref{fig3}. The fitted scaling $\langle T_c \rangle ^{\ast} \sim \left( r_b^{\ast} \right) ^{-\beta}$ is provided, with $\beta = 1.20 \pm 0.01$ and $\beta = 2.43 \pm 0.02$ for 2d and 3d, respectively.}
	\label{fig4}
\end{figure*}

\textbf{Brownian walkers in higher dimensions.} The mechanism of dynamic redundancy presented here allows the continuous emergence of new \textit{active} trajectories from $\mathbf{x}_0$, so the target will be found in a finite time even for the case of non-recurrent random walk trajectories, so both $\langle T \rangle$ and $\langle T_c \rangle$ will be finite for Brownian particles in arbitrary dimensions. In Fig. \ref{fig4} we show the results for $\langle T_c \rangle$ for the cases $d=2$ and $d=3$ to check whether the properties discussed above for the 1d case remain valid. From there, we confirm that there is still an optimum in $\langle T_c \rangle$ as a function of $r_b$. Since for this case we do not have a simple exact expression for $S_{sw}(t)$, we cannot predict analytically the particular behavior observed in the limits $r_b \gg 4D/x_0^2$ and $r_b \ll 4D/x_0^2$. However, generalizing our intuitive arguments above for 1d we observe that the former still satisfies $\langle T \rangle \sim 1/\ln{N}$ for the case of $N$ fixed, and so the scalings $\langle T \rangle \sim 1/\ln{r_b}$ and $\langle T_c \rangle \sim \chi/ \left( \ln{N} \right) ^2$ derived for the 1d case should persist. On the other side, for $r_b \ll 4D/x_0^2$ the same reasoning above should lead to $\langle T \rangle \approx \langle T_c \rangle \sim r_b^{-d/2}$ (taking into account again the properties of a diffusive flux to an absorbing trap in $d$ dimensions \cite{redner2001}). 

From Fig. \ref{fig4} we observe that the convergence to the asymptotic regimes $\chi \ll 1$ and $\chi \gg 1$  becomes extremely slow, so it is not possible at practice to fully observe the agreements with the scalings above. Also, we observe that the scaling between $\langle T_c \rangle ^{\ast}$ and $\left( r_b^{\ast} \right) ^{-1}$ becomes now nontrivial. In the insets of Fig. \ref{fig4} we provide a power-law fit $\langle T_c \rangle ^{\ast} \sim \left( r_b^{\ast} \right) ^{-\beta}$ to the results obtained, which predict exponents $\beta = 1.20 \pm 0.01$ (for 2d) and $\beta = 2.43 \pm 0.02$ (for 3d) for the region of situations explored. 

\textbf{Discussion and implications of our work.} Despite the interest raised by the redundancy principle, on one side, and stochastic resetting, on the other, in the literature on stochastic processes, the notion of dynamic redundancy introduced here represents a novel line of study which has remained unexplored to date. While the results above have been restricted to a very particular case for the sake of clarity, they show the potential interest that the problem has in terms of understanding and optimizing collective random searches, and they pave the way to explore more general situations. For instance, we have carried out some preliminary work with Brownian \textit{active} trajectories that can be terminated at a rate $r_d > 0$, and we observe that including the additional timescale $r_d^{-1}$ in the process makes that the MCST can reach minimum values for highly nontrivial combinations of $r_b$, $r_d$ (this will be explored in detail in a forthcoming publication). Also, the extension to non-Markovian birth-death mechanisms, or to more general movement processes (e.g. persistent random-walks, Lévy flights,...) and/or different geometries, offers a vast range of possibilities to explore.

On the other side, it is obvious that the results obtained above provide direct theoretical predictions for $\langle T \rangle$ and $\langle T_c \rangle$ that could be experimentally tested in real scenarios, e.g. in group animal foraging. For such cases, it is reasonable to ask whether the recruitment of new foragers can significantly reduce the search time, and/or increase the average energy intake for the group. Regulation of $n(t)$ then seems to be a simple mechanism that groups could use in order to orchestrate their searches and minimize their search efforts, and one could hypothesize this should actually play a role in the biological evolution and adaptation of such groups to their specific environmental conditions. These prospective ideas provide just some cues about the interest that the concept of dynamic redundancy could potentially have.

{\bf Acknowledgements}
DC acknowledges the organizers of the program in 'Mathematics of movement: an interdisciplinary approach to mutual challenges in animal ecology and cell biology' at the Isaac Newton Institute in Cambridge for fruitful discussions during his stay there. The Authors acknowledge the financial support of the Spanish government under grant PID2021-122893NB-C22.

\newpage

\section*{Dynamic redundancy as a mechanism to optimize collective random searches \\ SUPPLEMENTARY INFORMATION FILE}




\section*{Derivation of the survival probability $S(t)$}

In the main text we have already introduced the notation $S_N(t)$ for the survival probability of the target up to time $t$, conditioned to $n(t)=N$; this is, we divide all realizations of the stochastic redundancy process into those which have the same value of the birth-death process $n(t)$ at a particular time.

According to this, the explicit expression of $S_N(t)$ by definition is

\begin{eqnarray}
\nonumber    S_N(t) &=& \int_0^t dt_{N-1} \int_0^{t_{N-1}} dt_{N-2} \ldots \int_0^{t_2} dt_1  S_1 (t) S_1 (t-t_1) \ldots S_1 (t-t_{N-1})  = \\
    &=& \int_0^t dt_{N-1} \int_0^{t_{N-1}} dt_{N-2} \ldots \int_0^{t_2} dt_1 \left[ S_{sw}(t) e^{-r_b t_1} \right] \left[ S_{sw}(t-t_1) r_b e^{-r_b (t_2-t_1)} \right] \ldots \left[ S_{sw}(t-t_{N-1}) r_b e^{-r_b (t-t_{N-1})} \right],
    \label{survivalN1b}
\end{eqnarray}
where we have used $S_1(t)=S_{sw}(t)e^{-r_b t}$ in the last equality. The variables $t_i$ represent the random times at which new walkers become \textit{active} and so the process $n(t)$ increases by one unit. Then, $r_b e^{-r_b (t_{i}-t_{i-1})}$ represents the probability that the birth process requires a time $(t_{i}-t_{i-1})$ to increase from $n(t)=i-1$ to $n(t)=i$. 

Thanks to the Markovian nature of the birth-death process, the expression (\ref{survivalN1b}) conveniently simplifies to

\begin{equation}
    S_N(t)= r_b^{N-1} e^{-r_b t} \int_0^t dt_{N-1} \int_0^{t_{N-1}} dt_{N-2} \ldots \int_0^{t_2} dt_1 S_{sw}(t)  S_{sw}(t-t_1)  \ldots S_{sw}(t-t_{N-1}).
    \label{survivalN2b}
\end{equation}

On the other side, instead of considering all the random times $t_i$ explicitly, we can divide the whole process into two periods: (i) the interval $(0,t_1)$ during which $n(t)=1$, and (ii) the rest of the process. This leads to the recurrent expression

\begin{equation}
    S_N(t)= \int_0^t r_b S_1(t) S_{N-1}(t-t_1) dt_1,
    \label{survivalrec1}
\end{equation}
where the factor $r_b$ in the integral represents the rate at which the process switches from period (i) to period (ii).

From this expression, now it is immediate to check that the recurrent relation mentioned in the main text

\begin{equation}
    S_N(t)= r_b e^{-r_b t} S_{sw}(t) \int_0^t  S_{N-1}(t-t_1) dt_1,
    \label{survivalrec2}
\end{equation}
holds, and it is also immediate to verify that the general survival probability of the overall search process reads

\begin{equation}
    S(t) = \sum_{N=1}^{\infty} S_N(t) = S_{sw}(t) e^{-r_b (t-g(t))}.
    \label{survivaldef2}
\end{equation}
with $g(t)=\int_0^t S_{sw}(t)dt$.

Also, the fastest-first-passage time distribution through the target follows by deriving this expression with respect to $t$:

\begin{equation}
    f(t)= -\frac{dS(t)}{dt} = \left( f_{sw}(t) + r_b S_{sw}(t) [ 1-S_{sw}(t) ] \right) e^{-r_b (t-g(t))}
    \label{fptdef}
\end{equation}
where we define $f_{sw}(t)=-dS_{sw}/dt$ as the first-passage time distribution for the classical (single-walker) case.

Alternatively, the fastest-first-passage time distribution could be also derived by using recurrent arguments as above. So that, the recurrent relation can be written

\begin{equation}
    f_N(t)= \int_0^t r_b f_1(t) S_{N-1}(t-t_1) dt_1 + \int_0^t r_b S_1(t) f_{N-1}(t-t_1) dt_1,
    \label{fptrec1}
\end{equation}
for the contribution to the fastest-first-passage coming from those realizations where $n(t)=N$. Here, the first term on the right-hand side correspond to the case where it is the first \textit{active} walker (the one appearing at $t=0$) the one to reach the target first. The second term then corresponds to the case where any other of the $N-1$ \textit{active} walkers present at time $t$ reaches the target first. So that, taking into account that $f_1(t) = f_{sw} e^{-r_b t}$ and working out the recurrent relation, one finally obtains 

\begin{eqnarray}
    f_N(t)= r_b^{N-1} e^{-r_b t} \left[  f_{sw}(t) \frac{ (r_b g(t))^{N-1}}{(N-1)!}  +  S_{sw}(t) (1-S_{sw}(t))  \frac{ (r_b g(t))^{N-2}}{(N-2)!}  \right],
\end{eqnarray}

and finally using that $f(t)=\sum_{N=1}^{\infty} f_N(t)$ one recovers the expression in (\ref{fptdef}).

\section*{Derivation of the collective mean search time $\langle T_c \rangle$}

Again, we will consider by separate the realizations of the search process depending on the specific value of the birth-death process, $n(t)=N$, at the time at which the target is reached. By definition, if there are $N$ walkers \textit{active}, then the contribution of the first \active{walker} to the collective search time is $t$, and the contribution from any other \textit{active} walker will be $(t-t_i)$ (where $t_i$ represents the time at which tihs walker became \textit{active}). So that, the collective search time, averaged over all realizations of the process that lead to the target detection at time $t$, is

\begin{equation}
 \overline{T_c^{(N)}}(T) = r_b^{N-1} e^{-r_b T} \int_0^t dt_{N-1} \int_0^{t_{N-1}} dt_{N-2} \ldots \int_0^{t_2} dt_1 [ T+ (T-t_1) + \ldots + (T-t_{N-1}) ] G_N (T),
    \label{intensityN}
\end{equation}

where we have defined

\begin{equation}
    G_N (t) \equiv f_{sw}(t) \prod_{j=i}^{N-1} S_{sw} (t-t_j) + S_{sw}(t) \sum_{i=1}^{N-1} f_{sw}(t-t_i) \prod_{j \neq i} S_{sw} (t-t_j).
    \label{auxiliarN}
\end{equation}

Here, the first term on the right-hand side of (\ref{auxiliarN}) corresponds to the case where the first \textit{active} walker reaches the target at time $t$ before any of the other walkers do, as $\prod_{j=i}^{N-1} S_{sw} (t-t_j)$ is the joint survival probability for all the other \textit{active} walkers appearing at times $t_1$, $t_2$, $t_3$,...,$t_{N-1}$. Similarly, $\prod_{j \neq i} S_{sw} (t-t_j)$ represents the joint survival probability for all the \textit{active} walkers except the one appearing at time $t_j$; then, the last term in (\ref{auxiliarN}) gives the contribution from the realizations where the first walker to reach the target is the one becoming \textit{active} at time $t_i$.

In order to find a simple expression for $\overline{T_c^{(N)}}(T)$ we need to introduce now an additional function. So, we define $I_{N}(t)$ as the mean value of the collective search time by time $t$ provided that the target has not been reached yet by then. This satisfies equations analogous to (\ref{intensityN}-\ref{auxiliarN}) which read

\begin{equation}
 I_{N}(t) = r_b^{N-1} e^{-r_b t} \int_0^t dt_{N-1} \int_0^{t_{N-1}} dt_{N-2} \ldots \int_0^{t_2} dt_1 [ t+ (t-t_1) + \ldots + (t-t_{N-1}) ] H_N (t),
    \label{intensityN2}
\end{equation}

\begin{equation}
    H_N (t) \equiv S_{sw}(t) \prod_{j=i}^{N-1} S_{sw} (t-t_j) + S_{sw}(t) \sum_{i=1}^{N-1} S_{sw}(t-t_i) \prod_{j \neq i} S_{sw} (t-t_j).
    \label{auxiliarI}
\end{equation}

By its definition, the function $I_{N}(t)$ can be seen to satisfy the recurrent relation

\begin{equation}
    I_{N}(t) = r_b e^{-r_b t} S_{sw}(t) \int_0^t dt_1 \left( \frac{t-t_1}{t} I_{N-1}(t-t_1)+ S_{N-1}(t-t_1) \right),
\end{equation}

so applying this recurrence from $N=1$ on, it is possible to find a closed expression for it:

 \begin{equation}
      I_{N}(t) = e^{-r_b t} S_{sw} (t) \left[ \frac{ t \left(  r_b g(t) \right)^{N-1} }{(N-1)!}  + \frac{ r_b h(t) \left(  r_b g(t) \right)^{N-2}}{(N-2)!}      \right],  
 \end{equation}
where $h(t) \equiv \int_0^t tS_{sw}(t) dt$.  

With the help of the auxiliary function $I_{N}(t)$, we can write now a recurrence relation for $\overline{T_c^{(N)}}(T)$ which takes the form

\begin{eqnarray}
\nonumber    \overline{T_c^{(N)}}(T) &=& r_b e^{-r_b T} S_{sw}(T) \int_0^T dt_1 \left[ (T-t_1) \overline{T_c^{(N-1)}}(T-t_1)+ f_{N-1}(T-t_1) \right] \\
&+& r_b e^{-r_b T} f_{sw}(T) \int_0^T dt_1 \left[ S_{(N-1)}(T-t_1) + (T-t_1) I_{N-1}(T-t_1) \right].
\end{eqnarray}

Now, since we do have explicit expressions for $S_N (t)$, $f_N (t)$ and $I_{N}(t)$, it is also possible to reach recurrently a closed expression for $\overline{T_c^{(N)}}(T)$
\begin{eqnarray}
\nonumber    \overline{T_c^{(N)}}(T) &=& e^{-r_b T} S_{sw}(T) \left[ \frac{r_b (T+g(T)-2TS_{sw}(T)) \left( r_b g(T) \right)^{N-2}}{(N-2)!} + \frac{ r_b^2 (1-S_{sw}(T)) h(T) \left( r_b g(T) \right) ^{N-3} }{(N-3)!}  \right] \\
&+& e^{-r_b T} f_{sw}(T) \left[ \frac{ T \left( r_b g(T) \right) ^{N-1} }{(N-1)!} + \frac{r_b h(T) \left( r_b g(T) \right) ^{N-2} }{(N-2)!}  \right]  ,
\end{eqnarray}

and from this we get the final form for the mean collective search time,

\begin{eqnarray}
\nonumber   \overline{ T_c } (T) = \sum_{N=1}^{\infty} \overline{T_c^{(N)}}(T) &=&  e^{-r_b (T-g(T))} S_{sw}(T) \left[ r_b \left( t + g(T) - 2 t S_{sw}(T) \right) + r_b^2 \left( 1 - S_{sw}(T) \right) h(T) \right] \\
    &+& e^{-r_b (T-g(T))} f_{sw}(T) \left[ T + r_b h(T) \right]  ,
    \label{intensitydef2}
\end{eqnarray}

so now the MCST would be the integral of this expression over the possible values of the MFFPT, this is,

\begin{equation}
    \langle T_c \rangle = \int_0^{\infty} \overline{ T_c}(T) dT.
\end{equation}

\section*{Limiting behaviour of the MFFPT}
The MFFPT in 1d can be found from \eqref{survivaldef} and the survival probability for one walker $S_{sw}(t)$ as 
\begin{equation}
\left\langle T\right\rangle =\int_{0}^{\infty}S_{sw}(t)e^{-r_{b}t}e^{r_{b}\int_{0}^{t}S_{sw}(t')dt'}dt.
\label{mfpt}
\end{equation}
Integrating by parts, it turns into
\begin{equation}
\left\langle T\right\rangle =I-\frac{1}{r_{b}}\label{mfpt2b}
\end{equation}
where 
\begin{equation}
I=\int_{0}^{\infty}e^{-r_{b}t}e^{r_{b}\int_{0}^{t}S_{sw}(t')dt'}dt.\label{eq:I}
\end{equation}
The integral in the exponent of the integrand can be explicitly found in terms of the error function erf($\cdot$)
\[
\int_{0}^{t}S_{sw}(t')dt'=-\frac{x_{0}^{2}}{D}\textrm{erfc}\left(\frac{x_{0}}{\sqrt{4Dt}}\right)+t\textrm{erfc}\left(\frac{x_{0}}{\sqrt{4Dt}}\right)+x_{0}\sqrt{\frac{t}{\pi D}}e^{-\frac{x_{0}^{2}}{4Dt}}.
\]
Defining the dimensionless parameter 
\begin{eqnarray}
   \chi\equiv\frac{x_{0}^{2}r_{b}}{4D} 
   \label{chid}
\end{eqnarray}
and
the new integration variable $y=\sqrt{\chi/r_{b}t}$, the integral
(\ref{eq:I}) now becomes
\begin{equation}
I=\frac{2\chi}{r_{b}}\int_{0}^{\infty}e^{\varphi (\chi,y)}dy,\label{eq:II}
\end{equation}
where $\varphi (\chi,y)=\chi F(y)-3\ln y$ with
\begin{eqnarray}
F(y)=-\left(2+\frac{1}{y^{2}}\right)\textrm{erfc}(y)+\frac{2}{y\sqrt{\pi}}e^{-y^{2}}.   
\label{f}
\end{eqnarray}
This suggests to write \eqref{mfpt} in the scaling form
\begin{eqnarray}
      \left\langle T\right\rangle =\frac{x_{0}^{2}}{2D}\phi(\chi)
      \label{mT}
\end{eqnarray}
where
\begin{eqnarray}
\phi(\chi) =\int_{0}^{\infty}e^{\varphi(\chi,y)}dy-\frac{1}{2\chi}.    
\label{phi}
\end{eqnarray}

\begin{figure*}[t]
	\centering
		\includegraphics[width=0.6\textwidth]{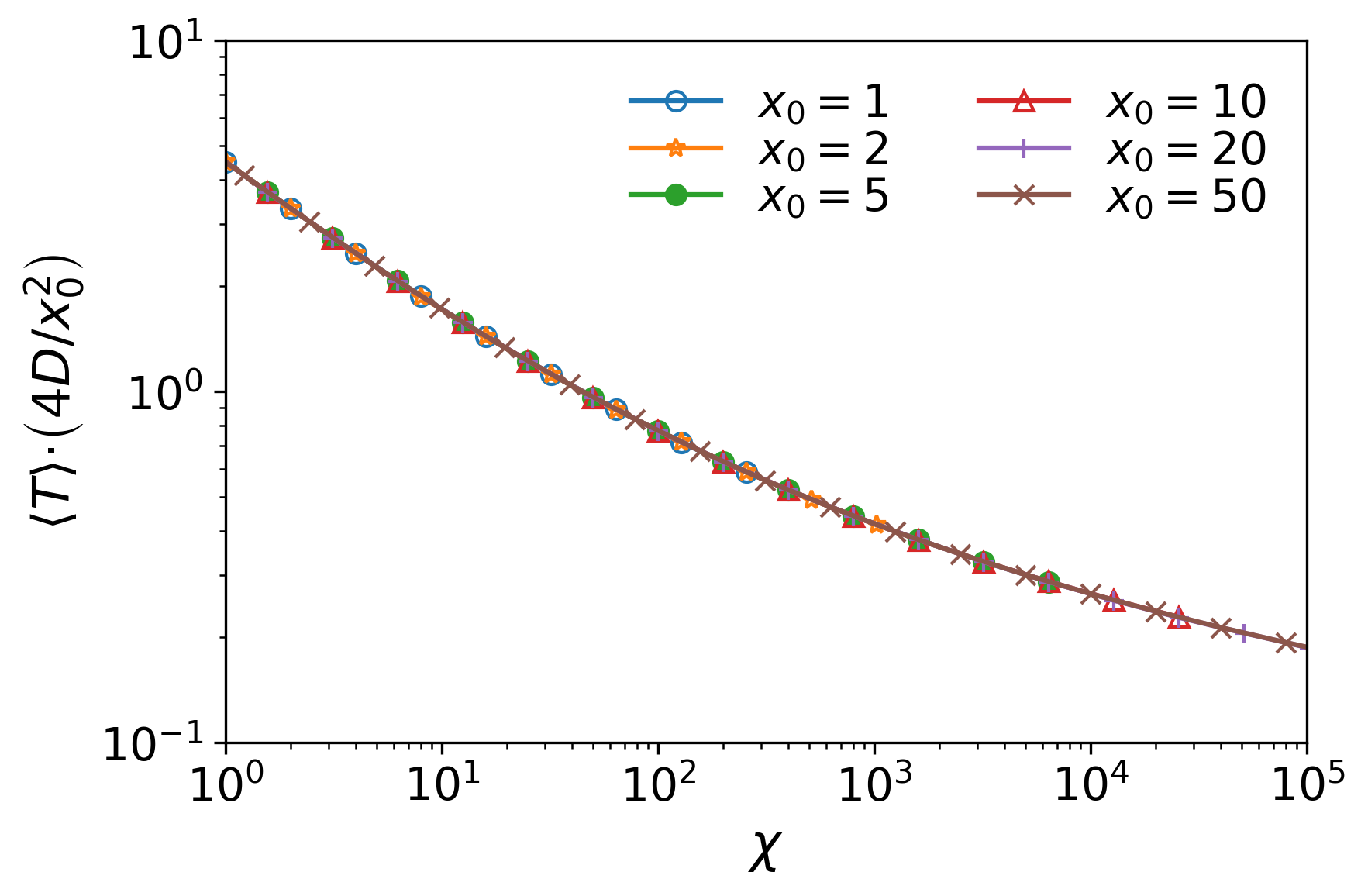}
		\caption{Nondimensional MFFPT as a function of the quotient between diffusive and birth scales, $\chi$, for different values of $x_0$ (see legend).}
	\label{fig:SI1}
\end{figure*}

The validity of this scaling is verified (for the case of Brownian walkers in 1d) by observing the collapse of the numerical results obtained for $\langle T \rangle$ when we plot them in the nondimensional form $4 D \langle T \rangle / x_0^2$ as a function of the variable $\chi$ (see Fig. \ref{fig:SI1}).

We note that the function $e^{\varphi(\chi,y)}$ is positive and has a global maximum 
at $y=y^{*}$ where $y^{*}$ satisfies $\varphi'(\chi,y^{*})=0$ with $\varphi''(\chi, y^{*})<0$ and the prime symbol means derivative with respect to $y$. Then, $y^*$ is solution of the transcendent equation
\begin{equation}
3y^{*2}=2\chi\textrm{erfc}(y^{*}),\label{eq:eq}
\end{equation}
i.e., $y^*$ is function of $\chi$ and it not difficult to show that $\partial y^*/\partial \chi >0$, so $y^*$ is a monotonically increasing function of $\chi$. 
This transcendent equation also lacks of analytical solution but we can find the scaling behaviour of $\chi$ with $y^*$ in the limits of small and large $\chi$. When $\chi$ is small, $y^*$ approaches to zero and $\textrm{erfc}(y^*)\simeq 1-2y^*/\sqrt{\pi}+O(y^{*3})$, so that $\chi \sim y^{*2}$. When $\chi$ is large, $y^*$ is also large and $\textrm{erfc}(y^*)\simeq e^{-y^{*2}}/y^*\sqrt{\pi}$, so that $\chi \sim y^{*3}e^{y^{*2}}$. 

The main contribution to the integral in Eq. \eqref{phi} comes from the vicinity of $e^{\varphi(\chi,y)}$ around $y=y^*$. To proceed further we consider two limiting situations, small $\chi$ and large $\chi$, separately.

\subsection*{$\chi \ll 1 $}

Let us consider the case $\chi \to 0$, this is, $r_{b}\ll4D/x_{0}^{2}$.  In this limit $y^*$ approaches $y=0$ as $\chi$ tends to zero, i.e. the function $\varphi (\chi,y)$ is peaked around $y=0$. To estimate the integral in Eq. \eqref{eq:II} we expand $\varphi (\chi,y)$ around $y=0$ and we find $\varphi (\chi,y)\simeq -\frac{1}{y^{2}}+\frac{4}{y\sqrt{\pi}}-3\ln(y)+O(0)$. Then
\begin{equation}
\int_{0}^{\infty}e^{\varphi(\chi, y)}dy\simeq\frac{1}{2\chi}+\frac{e^{4\chi/\pi}}{\sqrt{\chi}}\left[\textrm{erf}\left(2\sqrt{\frac{\chi}{\pi}}\right)+1\right]\simeq\frac{1}{2\chi}+\frac{1}{\sqrt{\chi}}+O(0)\label{eq:Iap1}
\end{equation}
where we have expanded the last equality for small $\chi$. Finally, from \eqref{phi} and (\ref{eq:Iap1})
$$
\phi(\chi)\simeq \frac{1}{\sqrt{\chi}}
$$
and from (\ref{mT})  
\[
\left\langle T\right\rangle \simeq\frac{2\sqrt{\chi}}{r_{b}}=\frac{x_{0}}{\sqrt{Dr_{b}}}.
\label{scaleTchismall}
\]

\subsection*{$\chi \gg 1 $}
Next we consider the opposite limit $\chi\to\infty$, i.e.,
$r_{b}\gg4D/x_{0}^{2}$. In this case, $y^{*}$ is such that $\int_{0}^{y^{*}}e^{\varphi(\chi,y)}dy \ll \int_{y^{*}}^{\infty}e^{\varphi(\chi,y)}dy$. 
Since the main contribution to the integral comes in the vicinity
of $y=y^{*}$ and $y^{*}$ is large, we approximate $\varphi(\chi,y)$
in the large $y$ limit, where $F(y)\simeq-y^{-5}e^{-y^{2}}/\sqrt{\pi}$.
Then,
\begin{eqnarray}
 \int_{0}^{\infty}e^{\varphi(\chi,y)}dy=\int_{0}^{\infty}\frac{e^{\chi F(y)}}{y^{3}}dy\simeq\int_{0}^{\infty}\frac{e^{-\frac{\chi}{\sqrt{\pi}}\frac{e^{-y^{2}}}{y^{5}}}}{y^{3}}dy\simeq\int_{y^{*}}^{\infty}\frac{e^{-\frac{\chi}{\sqrt{\pi}}\frac{e^{-y^{2}}}{y^{5}}}}{y^{3}}dy\sim\int_{y^{*}}^{\infty}\frac{dy}{y^{3}}\sim\frac{\text{1}}{y^{*2}}  
 \label{intex}
\end{eqnarray}
where we have considered $e^{-\frac{\chi}{\sqrt{\pi}}\frac{e^{-y^{2}}}{y^{5}}}\simeq 1$, since along the integration region $y$ is large. To find $y^{*}$ we solve (\ref{eq:eq}) for $\chi$ and expand for
large $y^{*}$ 
\[
\chi\simeq\frac{3\sqrt{\pi}}{2}y^{*3}e^{y^{*2}}+O(1/y^{*}).
\]
Solving this equation for $y^{*2}$ in terms of $\chi$ we find 
\begin{equation}
y^{*2}\simeq\frac{3}{2}W_{0}\left(\frac{2}{3}\left[\frac{2\chi}{3\sqrt{\pi}}\right]^{2/3}\right),
\label{eq:w0}
\end{equation}
where $W_{0}(x)$ is the principal branch of the Lambert function and can be approximated as $W_{0}(x)\simeq\ln(x)$ for large $x.$
Then, (\ref{eq:w0}) can be approximated through
\begin{equation}
y^{*2}\simeq\frac{3}{2}\ln\left(\frac{2}{3}\left[\frac{2\chi}{3\sqrt{\pi}}\right]^{2/3}\right)=\ln\left[\left(\frac{2}{3}\right)^{5/2}\frac{\chi}{\sqrt{\pi}}\right]\sim\ln\chi\quad\textrm{for}\quad\chi\to\infty.
\label{eq:yas}
\end{equation}
Additionally, from \eqref{intex} and \eqref{eq:yas} one has
\begin{equation}
\int_{0}^{\infty}e^{\varphi(\chi,y)}dy=\int_{0}^{\infty}\frac{e^{\chi F(y)}}{y^{3}}dy\sim\frac{1}{\ln\chi},
\label{eq:I2}
\end{equation}
and then combining \eqref{phi} and \eqref{eq:I2} one obtains the scaling
\begin{eqnarray}
\phi(\chi)\sim \frac{1}{\ln\chi}\quad\textrm{for}\quad\chi\to\infty.
    \label{fiapx}
\end{eqnarray}

\section*{Limiting behaviour of the MCST}
The integrals in the definitions of $g(t)$ and $h(t)$ can be explicitly written to express Eqs. (5-6) of the main text in terms of $T$. Introducing again the new variable $y\equiv \sqrt{\chi/r_bT}$ the mean collective search time can be written in the scaling form
\begin{eqnarray}
    \left\langle T_{c}\right\rangle =\frac{x_{0}^{2}}{2D}\phi_c(\chi),
    \label{Tc}
\end{eqnarray}
with
\begin{equation}
   \phi_{c}(\chi)=\int_{0}^{\infty}e^{\varphi_{c}(\chi,y)}dy
   \label{fic}
\end{equation}
and
$$
\varphi_{c}(\chi,y)=\chi F(y)-3\ln y+\ln\Phi(\chi,y).
$$

\begin{figure*}[t]
	\centering
		\includegraphics[width=0.6\textwidth]{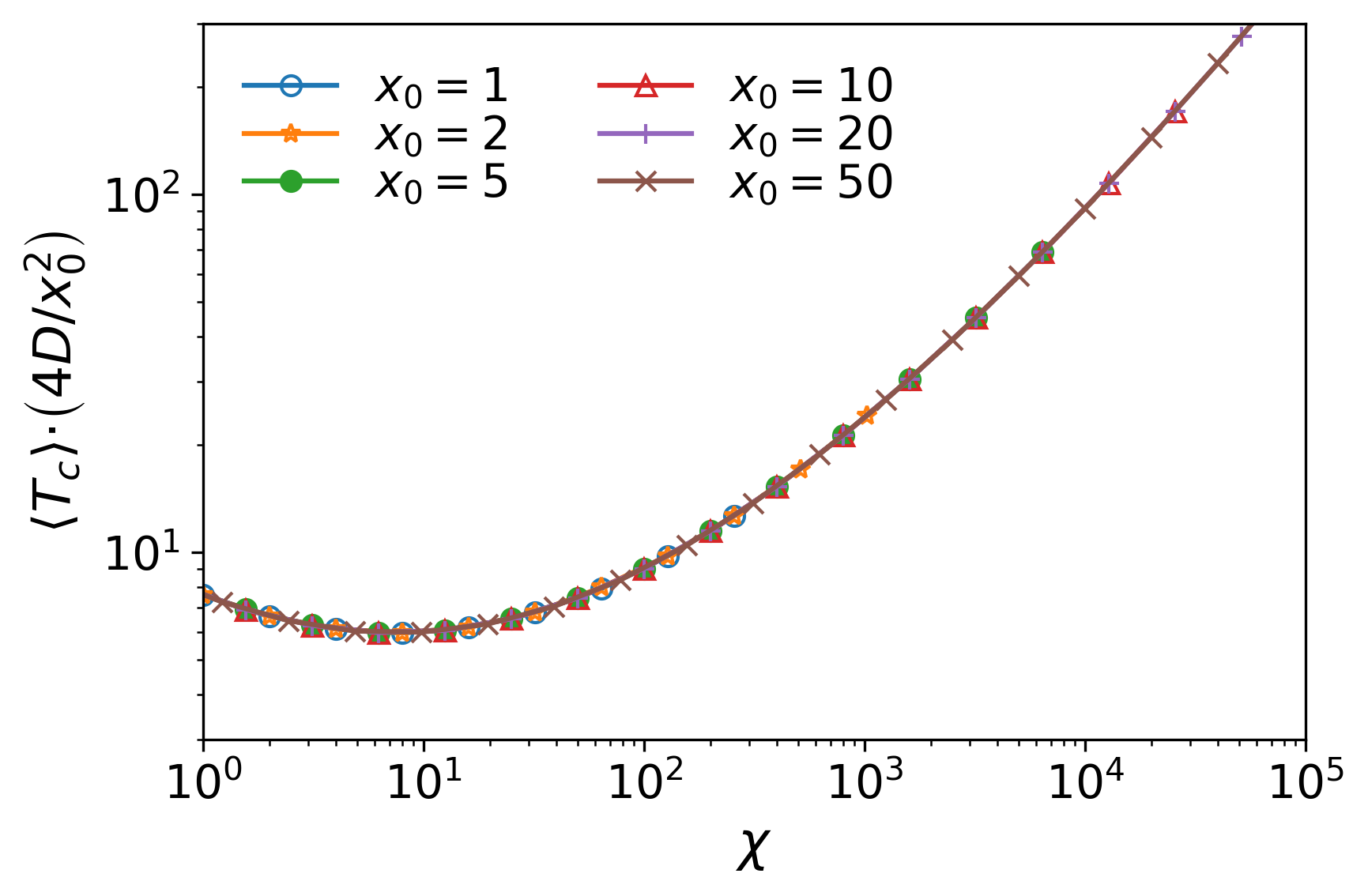}
		\caption{Nondimensional MCST as a function of the quotient between diffusive and birth scales, $\chi$, for different values of $x_0$ (see legend).}
	\label{fig:SI2}
\end{figure*}

Again, we verify the validity of the scaling proposed (for the case of Brownian motion in 1d) by plotting $4 D \langle T \rangle / x_0^2$ as a function of $\chi$ and checking how all curves obtained numerically collapse (Fig. \ref{fig:SI2}).

Here, $F(y)$ is defined in \eqref{f} and $\Phi(\chi,y)=\phi_{0}(y)+\chi\phi_{1}(y)+\chi^{2}\phi_{2}(y)$, with
\begin{eqnarray*}
   & & \phi_{0}(y)=\frac{y}{\sqrt{\pi}}e^{-y^{2}}\\
    & &\phi_{1}(y)=-(2y^{2}-1)\left[\frac{\textrm{erf}(y)\textrm{erfc}(y)}{y^{2}}+\frac{e^{-2y^{2}}}{3\pi}\right]+\frac{2y^{3}e^{-y^{2}}}{3\sqrt{\pi}}\textrm{erfc}(y)+\frac{5e^{-y^{2}}}{2y\sqrt{\pi}}\textrm{erf}(y)\\
  & &  \phi_{2}(y)=\textrm{erf}(y)\textrm{erfc}(y)\left[-\frac{2y^{2}-1}{3y^{3}\sqrt{\pi}}e^{-y^{2}}+\frac{\textrm{erf}(y)}{2y^{4}}+\frac{2}{3}\textrm{erfc}(y)\right].
\end{eqnarray*}
The function $e^{\varphi_{c}(\chi,y)}$ has a global maximum at $y=y^*$ where $y^*$ is solution to the equation $\varphi_{c}'(\chi,y^*)=0$, this is
\begin{eqnarray}
 y^{*}\Phi(y^{*},\chi)\chi f'(y^{*})+y^{*}\Phi'(y^{*},\chi)=3\Phi(y^{*},\chi),
 \label{eqt}
\end{eqnarray}
where the prime symbol means derivative with respect to $y^*$. Since $\varphi_{c}(\chi,y)$ increases monotonically with $\chi$, the position of the maximum $y^*$ moves with $\chi$ as for MFFPT. This transcendent equation also lacks of analytical solution but we can find the behaviour of $\chi$ with $y^*$ in the limits of small and large $\chi$. Inserting the definitions of $F(y)$ and $\Phi(y,\chi)$ into Eq. \eqref{eqt} we find, for large $\chi$, that
\begin{eqnarray}
    \chi \sim y^{*5}e^{y^{*2}}.
    \label{sc1}
\end{eqnarray}

\subsection*{$\chi \ll 1$}
In this limit $y^*$ tends to zero, i.e., the main contribution of the integral \eqref{fic} comes from the vicinity of $e^{\varphi_{c}(\chi,y)}$ near $y=0$. Expanding $F(y)$ and $\Phi (\chi, y)$ around $y=0$ we find
$$
F(y)\simeq-\frac{1}{y^{2}}+\frac{4}{y\sqrt{\pi}}+O(1),\quad\Phi(\chi,y)\simeq\frac{y}{\sqrt{\pi}}+\frac{2\chi}{y\sqrt{\pi}}+\frac{8\chi^{2}}{3\pi y^{2}}
$$
where we have expanded $\phi_0(y)$, $\phi_1(y)$ and $\phi_2(y)$ to the lowest order in $y$. Plugging these results into Eq. \eqref{fic} we find
$$
\int_{0}^{\infty}e^{\varphi_{c}(\chi,y)}dy\simeq\int_{0}^{\infty}e^{\frac{4u}{\sqrt{\pi}}-\frac{u^{2}}{\chi}}\left(\frac{1}{\chi\sqrt{\pi}}+\frac{2u^{2}}{\chi^{2}\sqrt{\pi}}+\frac{8u^{2}}{3\pi\chi^{2}}\right)dy=\frac{1}{\sqrt{\chi}}+O(1)
$$
where we have introduced the new variable $u=\chi /y$. Finally, from this result together with \eqref{fic} and \eqref{Tc} we get
$$
\left\langle T_c\right\rangle \simeq \frac{x_{0}}{\sqrt{Dr_{b}}},
$$
and so the result $\langle T \rangle \approx \langle T_c \rangle$ is recovered in this limit (see Eq. (\ref{scaleTchismall})), as discussed in the main text.

\subsection*{$\chi \gg 1$}
First we rewrite Eq. \eqref{fic} in the form
\begin{eqnarray}
    \phi_{c}(\chi)=\int_{0}^{\infty}\Phi(\chi,y)\frac{e^{\chi F(y)}}{y^{3}}dy
    \label{fic2}
\end{eqnarray}
and note that when $\chi$ is large, then $y^*$ is also large, as we argued below Eq. \eqref{eqt}. Since in this limit, the main contribution to the integral in \eqref{fic2} is due to the vicinity of $y^*$ and it is large, we expand $F(y)$ and $\Phi (\chi,y)$ for large $y$. In the large $y$ limit $F(y)\simeq-y^{-5}e^{-y^{2}}/\sqrt{\pi}$ and expanding $\phi_0(y)$, $\phi_1(y)$ and $\phi_2(y)$ we have
$$
\Phi(\chi,y)\simeq\frac{e^{-y^{2}}}{\sqrt{\pi}}\left(y+\frac{\chi}{2y}+\frac{\chi^{2}}{2y^{5}}\right).
$$
According to \eqref{sc1}, the dominant term of the above expression is the third term of the right hand side in the vicinity of $y^*$, so that
$$
\Phi(\chi,y)\simeq\frac{\chi^{2}}{2y^{5}}\frac{e^{-y^{2}}}{\sqrt{\pi}}.
$$
Hence, from \eqref{fic2}
$$
\phi_{c}(\chi)\simeq\frac{\chi ^2}{2}\int_{0}^{\infty}\frac{e^{-y^{2}}}{\sqrt{\pi}y^{5}}\frac{e^{-\frac{\chi}{\sqrt{\pi}}\frac{e^{-y^{2}}}{y^{5}}}}{y^{3}}dy=-\frac{\chi^{2}}{2}\frac{\partial}{\partial\chi}\int_{0}^{\infty}\frac{e^{-\frac{\chi}{\sqrt{\pi}}\frac{e^{-y^{2}}}{y^{5}}}}{y^{3}}dy,
$$
but the last integral was estimated in \eqref{eq:I2} so that from \eqref{Tc}
$$
\langle T_c \rangle \sim \phi_c(\chi)\sim -\chi^{2}\frac{\partial}{\partial\chi}\left(\frac{1}{\ln\chi}\right)=\frac{\chi}{(\ln\chi)^{2}}.
$$

\section*{optimal MCST}
The optimal value of $\left\langle T_{c}\right\rangle $ can be obtained by solving the equation $\partial \left\langle T_{c}\right\rangle /\partial \chi =0$ for $\chi$. Taking the derivative of \eqref{Tc} with respect to $\chi$ leads us to the transcendental equation
\begin{eqnarray}
    \int_{0}^{\infty}\frac{F(y)\phi_{0}(y)+\phi_{1}(y)}{y^{3}}e^{\chi F(y)}dy+\chi\int_{0}^{\infty}\frac{F(y)\phi_{1}(y)+2\phi_{2}(y)}{y^{3}}e^{\chi F(y)}dy+\chi^{2}\int_{0}^{\infty}\frac{F(y)\phi_{2}(y)}{y^{3}}e^{\chi F(y)}dy=0,
    \label{eqchi}
\end{eqnarray}
which is an integral equation for $\chi$. Solving \eqref{eqchi} numerically one finds $\chi ^*=1.8877$, which from \eqref{chid} corresponds to the optimal birth rate $r_b^*\simeq 7.551 D/x_0^2$. Substituting $\chi=\chi^*$ in \eqref{Tc} one obtains $ \left\langle T_{c}\right\rangle ^*\simeq 1.489 x_0^2/D$.

\end{document}